\documentclass[aps,prb,floats,twocolumn,twoside,floatfix,superscriptaddress]{revtex4}
\usepackage{bm,amsmath,amssymb,amsthm} 
\usepackage[utf8]{inputenc} 
\usepackage{psfrag}
\usepackage{graphicx}
\usepackage{units} 
\usepackage{gensymb}  

\begin{document}
\title{Unconventional cycles, pseudoadiabatics and  multiple adiabatic points}
\author{Jeferson J. Arenzon}
\affiliation{Instituto de Física, Universidade Federal do Rio Grande do Sul, CP 15051, 91501-970, Porto Alegre,
RS, Brazil}
\date{\today}
\begin{abstract}
Unconventional cycles provide a useful didactic resource to discuss the second law of thermodynamics applied to thermal motors and their efficiency. In most cases they involve a negative slope, linear process that presents an adiabatic point where the process is tangent
to an adiabatic curve and $\delta Q=0$, signalling that the flow of heat is reversed.
We introduce a parabolic process, still simple enough to be fully explored analitically in
order to deal with the usual follow up question on the possibility of having more than
one adiabatic point. Having one (linear), two (parabolic) or more such points
allow the construction of reversible,
non-isoentropic processes, that we call pseudoadiabatics, whose total heat exchanged is zero.
\end{abstract}

\maketitle

\section{Introduction}

Along some of the idealized, conventional processes most often considered in thermodynamics, the exchanged heat $\delta Q$ is either constant (isocoric or isobaric), null (adiabatic) or monotonously varying with the volume (isothermal or positively sloped linear processes).
Importantly, in all these cases, $\delta Q$ is either positive or negative along the whole process.
Nonetheless, in the so-called unconventional (albeit idealized) cycles, $\delta Q$ has a smooth, continuous reversal of direction (signal)~\cite{WiKi80,DiMo94,Valentine95,Hernandez95,Leff95,KaMaSh96,KaSh97,Bucher99}, changing from endo to exothermic.
In this so-called adiabatic point,  as the name indicates, the process $P(V)$ is tangent to an adiabatic curve~\cite{DiMo94} and $\delta Q=0$. 
The simplest, most discussed example is the linear, negative slope process~\cite{WiKi80,DiMo94,Valentine95,Hernandez95,KaMaSh96,Leff95,Bucher99} while 
more complex cases involve circular~\cite{DiMo94,MaSh00,LiMoPeRu03}, 
elliptical~\cite{VeGoWhHe02} and multilobed~\cite{Chen07} cycles. 
Although the linear case may only have a single~\cite{DiMo94} adiabatic point (because adiabatic curves cannot cross), there are numerical results showing that multiple adiabatic points are possible on the circular cycle~\cite{LiMoPeRu03}. Our main objective here is to present a simple, treatable example that can be discussed on a basic course of thermodynamics where multiple adiabatic points may appear, the parabolic process. Differently from the linear
process, the parabolic one with $Q=0$ can be visually disguised as an adiabatic process
and thus become a good starting point to discuss the related subtleties of these
processes and the cycles they may belong to.
Before that, we revisit the linear case and explore the possibility of having a 
pseudoadiabatic process, a reversible, non 
isoentropic process with zero net exchanged heat ($Q=0$).


Consider one mol of a monoatomic ideal gas and the linear process
passing through $(V_0,P_0)$ and $(2V_0,\alpha P_0)$,
where $\alpha< 1$ is a control parameter. Introducing
adimensional variables, $p\equiv P/P_0$ and $v\equiv V/V_0$, 
\begin{equation}
p = (\alpha-1)(v-1)+1,
\label{eq.plinear}
\end{equation}
where $0\leq v\leq v_{\scriptscriptstyle\rm MAX}=(\alpha-2)/(\alpha-1)$. 
The adimensional, rescaled temperature, $t\equiv RT/P_0V_0=pv$, has
a maximum at the point T, where $v_{\scriptscriptstyle\rm T}=v_{\scriptscriptstyle\rm MAX}/2$.
Following Ref.~\onlinecite{DiMo94},
the heat exchanged during an infinitesimal step along the process is obtained from the first
law of thermodynamics, $\delta Q=c_{\scriptscriptstyle\rm V}dT+PdV$ (with $c_{\scriptscriptstyle\rm V}=3R/2$), and the equation of state $PV=RT$,
\begin{equation}
\delta Q=P_0V_0\left[ 4(\alpha-1)v+\frac{5}{2}(2-\alpha)\right] dv.
\label{eq.deltaQ}
\end{equation}
The point A where the heat flow is reversed, $\delta Q=0$, is
\begin{equation}
v_{\scriptscriptstyle\rm A} = \frac{5}{8}  v_{\scriptscriptstyle\rm MAX},
\end{equation}
thus occurring at a larger volume than T, i.e., $v_{\scriptscriptstyle\rm A}>v_{\scriptscriptstyle\rm T}$.
 Eliminating 
$\alpha$ we write $p_{\scriptscriptstyle\rm A}=3v_{\scriptscriptstyle\rm A}/(8v_{\scriptscriptstyle\rm A}-5)$, locating all adiabatic points for the chosen parametrization (equivalently, $p_{\scriptscriptstyle\rm T}=v_{\scriptscriptstyle\rm T}/(2v_{\scriptscriptstyle\rm T}-1)$ for T). By changing the declivity of the linear process, $v_{\scriptscriptstyle\rm A}$ goes from 5/8 ($\alpha\to -\infty$) to $\infty$ ($\alpha\to 1$). Moreover, the specific heat
along the linear process~\cite{Hernandez95},
\begin{equation}
c=\left(\frac{\partial Q}{\partial T}\right)_{\rm lin}= 2R\frac{v-v_{\scriptscriptstyle\rm A}}{v-v_{\scriptscriptstyle\rm T}},
\label{eq.specific1}
\end{equation}
is negative for $v_{\scriptscriptstyle\rm T}<v<v_{\scriptscriptstyle\rm A}$ (positive 
when $v<v_{\scriptscriptstyle\rm T}$ or $v>v_{\scriptscriptstyle\rm A}$), 
zero at A and divergent at T.

\begin{figure}[htb]
\includegraphics[width=8cm]{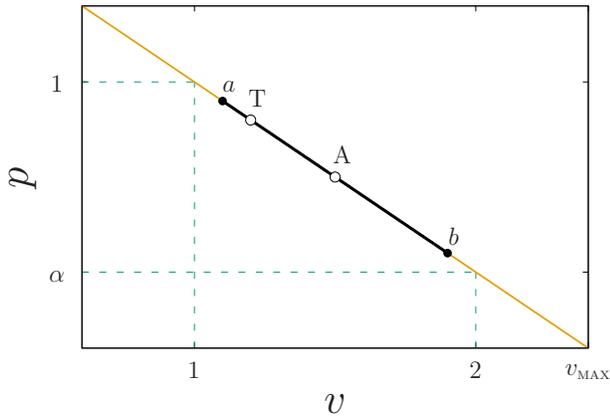}
\caption{Linear process, Eq.~(\ref{eq.plinear}), with $\alpha=2/7$ (such that
$v_{\scriptscriptstyle\rm A}=3/2$) and adimensional variables, $v\equiv V/V_0$ and
$p\equiv P/P_0$. The segment $\overline{ab}$, symmetric around A, is referred as 
a pseudoadiabatic process because 
$Q_{ab}=0$.  
The point T locates the maximum of temperature while A, where the heat flow is reversed ($\delta Q=0$), is tangent to an adiabatic curve (adiabatic point).}
\label{fig.linear}
\end{figure}

Particularly interesting is the case in which the total heat exchanged is
zero. This occurs when the points are symmetrically chosen around $v_{\scriptscriptstyle\rm A}$ (e.g., $a$ and $b$ in Fig.~\ref{fig.linear}). Indeed, integrating $\delta Q$ from $v_a$ to $v_b$ and imposing $Q_{ab}=0$, we get that $v_{\scriptscriptstyle\rm A}$ is the average
between $v_a$ and $v_b$:
\begin{equation}
v_a+v_b=\frac{5}{4} v_{\scriptscriptstyle\rm MAX} = 2v_{\scriptscriptstyle\rm A}.
\end{equation} 
This result is expected from Eq.~(\ref{eq.deltaQ}) whose dependence on the
volume is linear.
Such process, that we call pseudoadiabatic, offers a myriad of interesting problems to be explored on itself and
as part of thermodynamical cycles.
Not only to emphasize the differences with an actually 
adiabatic process, but also as a simple way to introduce those
unconventional cycles and discuss the subtleties on the evaluation of thermodynamical
efficiencies.


The possibility of existing several adiabatic points has attracted little attention, having being briefly mentioned for the circular process~\cite{LiMoPeRu03}, but is a usual question following the discussion of the linear process. Because the circular cycle usually resorts to numerical techniques, it is important to have an intermediate and treatable case. In the next section we thus consider the simplest case beyond the linear process, the parabolic expansion.

\section{The parabolic process}

\begin{figure}[htb]
\includegraphics[width=8cm]{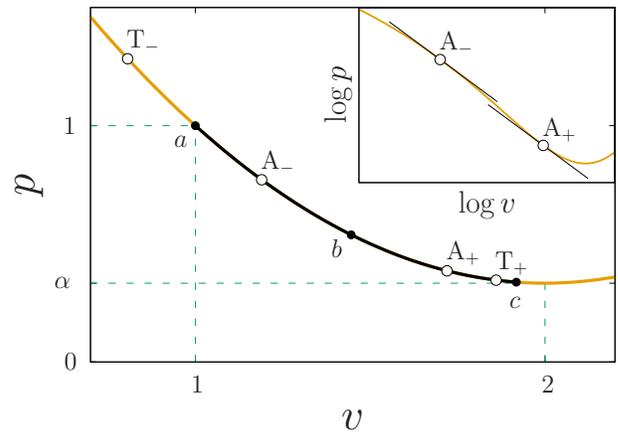}
\caption{Parabolic process, Eq.~(\ref{eq.pv}), for $\alpha=1/3$ with the two adiabatic
points, A$_-$ and A$_+$.
Also shown are the two extremes of temperature T$_+$ and T$_-$ that, as in the linear
case, do not coincide with the adiabatic points. 
The points $a$, $b$ and $c$ mark the three possible pseudoadiabatic processes starting
at $v_a=1$: $Q_{ab}=Q_{bc}=Q_{ac}=0$. Inset: log-log plot showing the two adiabatics (straight, parallel lines with declivity $\gamma\equiv c_{\scriptscriptstyle\rm P}/c_{\scriptscriptstyle\rm V}=5/3$ for the monoatomic ideal gas considered here)
tangent to the parabola.}
\label{fig.pv}
\end{figure}

\begin{figure}[htb]
\includegraphics[width=8cm]{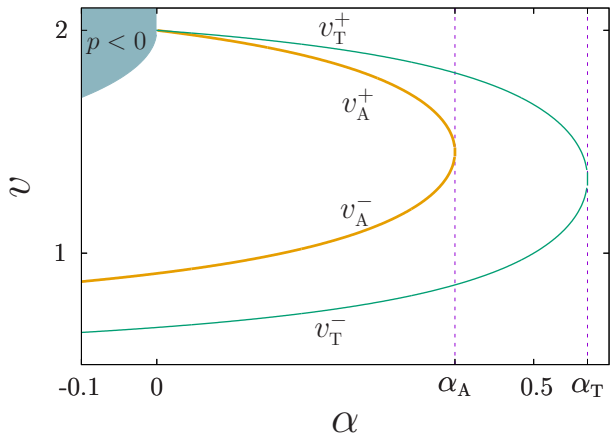}
\caption{Volume $v_{\scriptscriptstyle\rm A}^{\pm}$ of the adiabatic points A$_+$ and A$_-$, as a function of $\alpha$, for the parabolic process, Eq.~(\ref{eq.pv}), along with the minimum (T$_+$) and maximum (T$_-$) of temperature. The shaded region appearing for $\alpha<0$ corresponds
to $p<0$.}
\label{fig.va}
\end{figure}

Consider again one mol of a monoatomic ideal gas and the parabolic process $P=a_1V^2+a_2V+a_3$. 
The (dimensional) coefficients $\{a_i\}$ are such that the parabola goes through the point $(V_0,P_0)$ and has its minimum 
at $(2V_0,\alpha P_0)$ where it is, by construction, parallel to an
isobaric process. Using the adimensional variables $p$ 
and $v$, we write
\begin{equation}
p=(1-\alpha)(v-1)(v-3)+1,
\label{eq.pv}
\end{equation}
where $\alpha$ remains as a free parameter. An example with $\alpha=1/3$ is shown in Fig.~\ref{fig.pv}. The rescaled temperature $t$  has extremes at two points, $T_+$ and $T_-$, a minimum and a maximum, respectively: 
\begin{equation}
v_{\scriptscriptstyle\rm T}^{\pm} = \frac{4}{3} \pm \frac{1}{3}\sqrt{\frac{4-7\alpha}{1-\alpha}}.
\end{equation}
At these points the curve $p(v)$ is tangent to different isothermal processes.
Between these two points, and generalizing the linear case, there may be two adiabatic points,
A$_+$ and A$_-$. 
By analogy with Eq.~(\ref{eq.deltaQ}), the heat exchanged along the parabolic 
process is $\delta Q=q(v)P_0V_0dv$ with
$$
q(v) =   \frac{11}{2}(1-\alpha)v^2-16(1-\alpha)v+10-\frac{15}{2}\alpha.
$$
The two possible adiabatic points are located at
\begin{equation}
v_{\scriptscriptstyle\rm A}^{\pm} = \frac{16}{11}\pm \frac{1}{11}\sqrt{\frac{36-91\alpha}{1-\alpha}},
\label{eq.parab.va}
\end{equation}
shown in Fig.~\ref{fig.va} along with $v_{\scriptscriptstyle\rm T}^{\pm}$.
In the interval 
$v_{\scriptscriptstyle\rm A}^-<v<v_{\scriptscriptstyle\rm A}^+$, $\delta Q<0$ and
heat flows out of the system. Thus, A$_-$ plays a role similar to A in the linear
process, separating regions, as $v$ increases, with $\delta Q>0$ and 
$\delta Q<0$, respectively. Crossing A$_+$, the order is reversed, from  negative to positive heat.  
For $\alpha<0$, the point A$_+$ is not present anymore, 
a region with $p<0$ develops, instead, above $v=2+\sqrt{\alpha/(\alpha-1)}$, 
shaded in Fig.~\ref{fig.va}. 
The A$_+$ and A$_-$ points merge at $\alpha_{\scriptscriptstyle\rm A}=36/91$, 
$v_{\scriptscriptstyle\rm A}^-=v_{\scriptscriptstyle\rm A}^+=16/11$
 and although one 
adiabatic point still exists, there is no change in the heat flow direction.
For $\alpha_{\scriptscriptstyle\rm A}<\alpha<\alpha_{\scriptscriptstyle\rm T}$,
albeit $\delta Q>0$ along the whole process, there are still extremes of temperature
at T$_+$ and T$_-$. At $\alpha=\alpha_{\scriptscriptstyle\rm T}$
these points merge and the temperature has an inflection point associated with a divergent
specific heat (see below). 
The region where both adiabatic points may be present is thus $0\leq\alpha<\alpha_{\scriptscriptstyle\rm A}$.
Of course, the presence or not of such points on a given process also
depends on the initial and final states chosen along the parabolic curve.

In the detail of Fig~\ref{fig.pv}, the parabolic process is plotted on a log-log scale. On such
a logarithm scale, both the isothermal (not shown) and adiabatic processes are seen as straight
lines (with 
declivity 1 and $\gamma$, respectively) and provide a graphical way
to locate the adiabatic points~\cite{ShKa14}. Indeed, the two adiabatics that are tangent to the 
parabola at A$_+$ and A$_-$ are clearly seen as straight, parallel lines. Between A$_-$ and A$_+$ 
the slope is smaller than the slope of an adiabatic, $(dp/dv)_{\scriptscriptstyle\rm S}=-\gamma p/v$ 
(or, equivalently, $(d\ln p/d\ln v)_{\scriptscriptstyle\rm S}=-\gamma$) and there is heat leaving
the system. Outside this interval, the slope is higher compared to the adiabatic and heat enters the system.

Pseudoadiabatic processes may be defined for the parabola as well, although they are
no longer symmetric around an adiabatic point. Moreover, given an initial
point $a$ to the left of A$_-$, there may be two points $b$ and $c$, 
such that $Q_{ab}=Q_{ac}=0$ (there is, of course, a third possible pseudoadiabatic process, $Q_{bc}=0$). Choosing, for example, 
$v_a=1<v_{\scriptscriptstyle\rm A}^-$, and
integrating $q(v)$ up to the point where the overall heat would be zero,
one gets two non trivial solutions, $v_{\scriptscriptstyle\rm A}^-<v_b<v_{\scriptscriptstyle\rm A}^+$ and $v_c>v_{\scriptscriptstyle\rm A}^+$. Fig.~\ref{fig.pv} shows
an example (thick, black line).

\begin{figure}[htb]
\includegraphics[width=8cm]{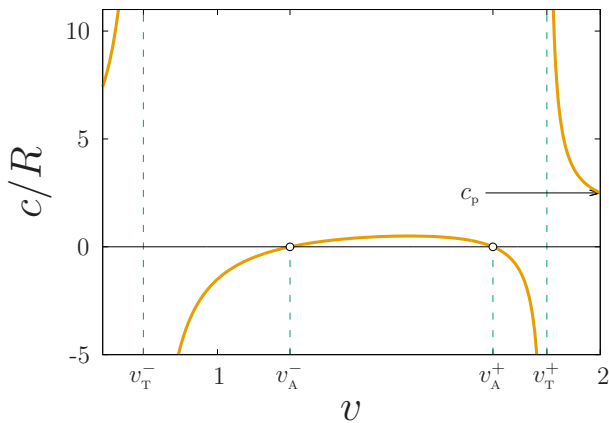}
\caption{Specific heat, Eq.~(\ref{eq.specific}), for the process of Fig.~\ref{fig.pv} with $\alpha=1/3$. 
There are divergences at T$_+$ and T$_-$ because the temperatures are extreme and $c$ may
be both negative ($v_{\scriptscriptstyle\rm T}^-<v<v_{\scriptscriptstyle\rm A}^-$ or 
$v_{\scriptscriptstyle\rm A}^+<v<v_{\scriptscriptstyle\rm T}^+$) or positive (otherwise). At
$v=2$, $c=c_{\scriptscriptstyle\rm P}=5R/2$.}
\label{fig.specific}
\end{figure}

The specific heat along the process may be obtained through $c\equiv (\partial Q/\partial T)_{\rm par}$,
generalizing the expression given in Eq.~(\ref{eq.specific1}) for the linear process:
\begin{equation}
c = \frac{R}{2}\frac{(v-v_{\scriptscriptstyle\rm A}^+)(v-v_{\scriptscriptstyle\rm A}^-)}{(v-v_{\scriptscriptstyle\rm T}^+)(v-v_{\scriptscriptstyle\rm T}^-)}.
\label{eq.specific}
\end{equation}
As illustrated in Fig.~\ref{fig.specific}, $c$ diverges at $v_{\scriptscriptstyle\rm T}^{\pm}$ where the parabola is tangent to isothermal processes and is
zero at $v_{\scriptscriptstyle\rm A}^{\pm}$ where it is tangent to the two adiabatic ones. 
Notice also that $v=2$ corresponds, by
construction, to the minimum of the parabolic process. Thus, because the infinitesimal
process is flat and parallel to an isobaric curve, $c=c_{\scriptscriptstyle\rm P}=5R/2$. 

\section{Discussion and conclusions}

Despite the name, unconventional cycles offer a good conceptual tool to discuss some subtleties 
related to thermodynamic cycles. For example, in the ``Saddly Cannot'' cycle introduced in Ref.~\onlinecite{WiKi80}, two points on an adiabatic curve are connected through a linear process similar to Eq.~(\ref{eq.plinear}).
When evaluating only the net exchanged heat in each process (zero and positive, respectively), one may erroneously conclude that this cycle violates the second law of thermodynamics, with full efficiency and no losses. By identifying the adiabatic point where $\delta Q=0$, it is possible to dismiss the apparent paradox, find the missing region with negative heat and correctly evaluate the efficiency. Another instance where recognizing those points is important is in less idealized cycles as they lack a clear transition between well known processes, and the round, broad crossovers make them more prone to exhibit adiabatic points, probably more than one. Although the linear,
negative slopped process may present only one, the
existence of processes with multiple adiabatic points had been numerically observed in 
complex cycles like the circular one. Thus, our objectives were twofold. First, to introduce 
a process with more than one adiabatic point, yet simple enough to be tackled analytically.
For that we considered a parabolic process that may present up to two such points. 
This leads to the second objective, the discussion of processes that, due to the
presence of at least one adiabatic point, have a zero total heat exchanged 
(at variance with an adiabatic process that has $\delta Q=0$ everywhere). We called
such processes as pseudoadiabatic, where the
heat exchanged at one side of an adiabatic point is compensated on the other side, 
the net heat exchanged being null.
A pseudoadiabatic parabolic process, not being straight, can be easily misrecognized for an actual adiabatic 
one and be a source of error and confusion when only the net heat
exchanged is evaluated. Addressing the differences between them can prove very useful for
the students.

More than two adiabatic points is also possible~\cite{LiMoPeRu03,Chen07}. If one considers only the
third quadrant of the circular cycle, it is possible to have up to three such points
in that region. By finding a curve that conveniently changes its curvature, it might be possible
to have even more. The parabolic process is interesting because we can obtain simple, closed expressions for it, thus being a useful didactic resource while discussing the subtleties of 
thermodynamical cycles and the respective efficiencies.



\begin{acknowledgments}
I thank Claudio Schneider for many interesting discussions on thermodynamics over the years
and for a critical reading of the manuscript. Also,  the INCT Sistemas Complexos and the Brazilian agencies CNPq, Fapergs and 
CAPES for partial financial support. 
\end{acknowledgments}


\end{document}